Decadal attribution of historic temperature and ocean heat content change to anthropogenic emissions

E.J.L. Larson[1,2] R.W. Portmann[1], S. Solomon[3], and D.M. Murphy[1]

[1]Earth System Research Laboratory, National Oceanic and Atmospheric Administration, Boulder, Colorado, USA
[2]Department of Organismic and Evolutionary Biology, Harvard University, Cambridge, MA 02138, USA
[3]Department of Earth, Atmospheric, and Planetary Sciences, Massachusetts Institute of Technology, Cambridge, MA 02139, USA

Corresponding author: Erik Larson (erik_larson@fas.harvard.edu)

**Abstract**

We present an alternative method of calculating the historical effective radiative forcing using the observed temperature record and a kernel based on the CMIP5 temperature response. This estimate is the effective radiative forcing time series that the average climate model would need to simulate the observed global mean surface temperature anomalies. We further infer the anthropogenic aerosols radiative forcing as a residual using the better-known greenhouse gas radiative forcing. This allows an independent estimate of anthropogenic aerosol radiative forcing, which suggests a cooling influence due to aerosols in the early part of the 20th century. The temporal kernels are also used to calculate decadal contributions from the dominant forcing agents to present day temperature, ocean heat content, and thermosteric sea level rise. The current global mean temperature anomaly is dominated by emissions in the past two decades, while current ocean heat content is more strongly affected by earlier decades.

**1 Introduction**

Earth's climate has changed markedly since the industrial revolution. The global mean two-meter air temperature (GMST) of Earth has risen about a degree Celsius since 1850 and is projected to rise much more (Collins et al., 2013; Karl et al., 2015; Kirtman et al. 2013). Associated with this warming are increased ocean heat content (OHC) and sea level rise (SLR). These climate changes are largely due to anthropogenic emissions of greenhouse gases (GHG), which are partially offset by natural and anthropogenic aerosol emissions (e.g., Bindoff et al., 2013; Charleson et al., 1991; Murphy et al., 2009).

GHGs and aerosols affect the climate by changing the radiative imbalance at the top of the atmosphere (TOA), also known as radiative forcing (RF). The radiative forcing and response cannot be distinguished by observations; therefore models are used to estimate it. There are several definitions of radiative forcing based on their method of calculation and each provides a slightly different value (Myhre et al., 2013; see also Forster and Taylor, 2006; Gregory et al., 2004; Hansen et al., 2005; Larson and Portmann, 2016; Shine, 2003). Here we calculate the effective radiative forcing (ERF), which includes rapid adjustments in the stratosphere, tropospheric adjustments of temperature, water vapor, and clouds, and some land surface temperature adjustment. We also note that throughout the paper we compare with and use other definitions of forcing under the assumption that they are not very different from the ERF (Hansen et al., 2005).

Climate models specify emissions scenarios and calculate variables (GMST, OHC, TOA radiation, etc) which are used to reconstruct the ERF (Gregory et al., 2004; Forster and Taylor, 2006; Larson and Portmann, 2016). This study reverses this direction and uses the observed GMST and a CMIP5 derived kernel to calculate ERF. The CMIP5 derived kernels have previously been used to diagnose forcing in transient climate simulations (Larson and Portmann, 2016). We apply the method to GMST observations that extend back to 1880 to produce an observationally based estimate of the historical ERF. These kernels allow us to combine multiple climate models, which increases the confidence of the results. While this estimate of the historical ERF is observationally based, it is still dependent on climate models and should be interpreted in the following way: This is the ERF time series that the average CMIP5 climate model would need to simulate the observed GMST anomalies. To the extent that these climate models represent the Earth system, this is the ERF that the Earth system experienced. Climate models with different sensitives (e.g. Gettelman et al., 2019) would yield different ERF.

Estimates of ERF are important for constraining the climate sensitivity of the Earth system. There is a large range of estimates from models and observations (Armour, 2017; Rohling et al., 2017; Storelvmo et al., 2017) and this uncertainty limits confidence in future projections from climate models. The uncertainty in the direct and indirect aerosol ERF is larger than any other forcing agent (Myhre et al., 2013). We calculate the anthropogenic aerosol ERF as a residual from our historical estimates and the better-known GHG and volcanic forcing.

Along with estimating the historical anthropogenic ERF, the temporal kernel method also allows for apportionment of the current GMST, OHC, and thermosteric SLR anomalies to past emissions. The CMIP5 models report thermosteric SLR, but not the total SLR. Warming due to $CO_2$ dominates the energy budget and is realized over long timescales (Myhre et al., 2013; Solomon et al., 2009). Furthermore, the Earth is committed to remain warm even if $CO_2$ emissions are halted (Archer, 2005; Solomon et al., 2009). A model incorporating both the atmospheric lifetimes of GHGs and aerosols and the climate response to changes in ERF from the emissions is used to apportion the current GMST and OHC anomalies to past decades of emissions.

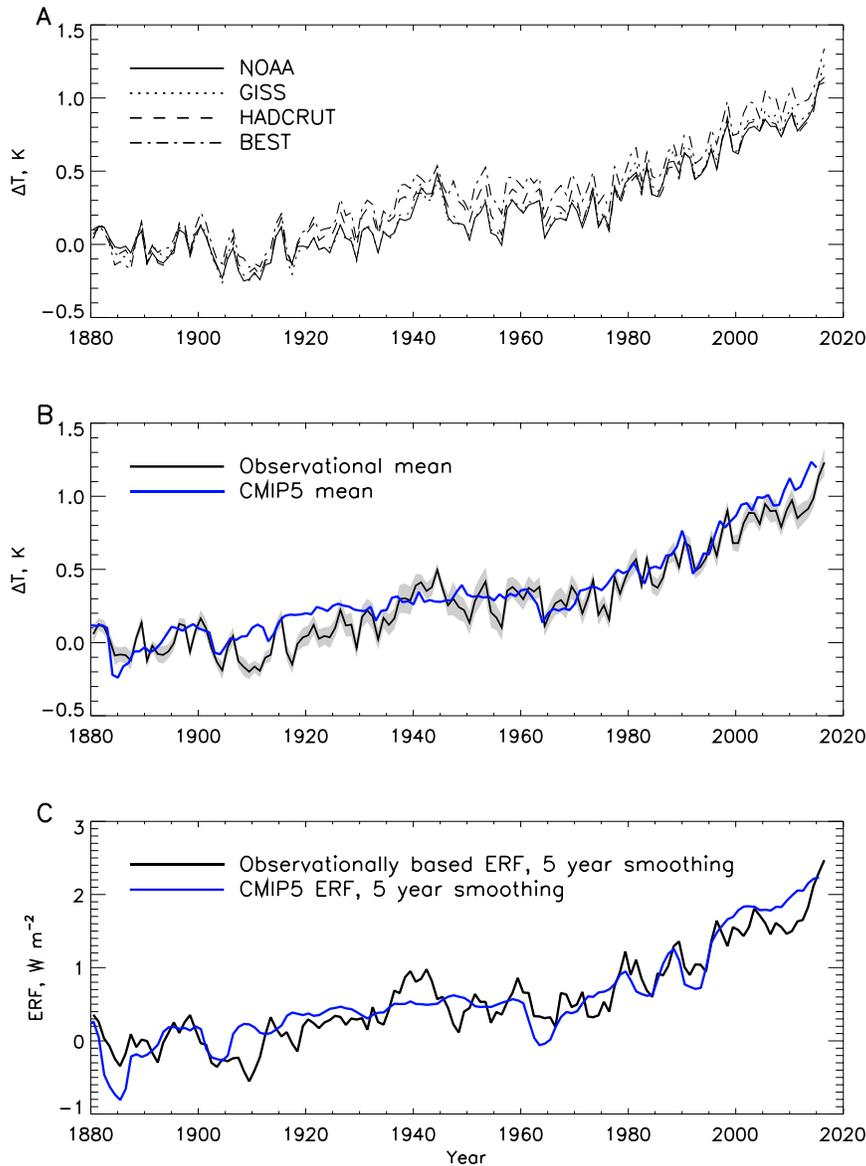

Figure 1. A) Observed annual average temperature anomaly relative to 1880-1900. B) The mean and standard deviation of the temperature time series along with the CMIP5 multi-model mean. C) The 5-year smoothed ERF calculated using the mean temperature from part B compared with the 5-year smoothed CMIP5 multi-model mean ERF.

## 2 Estimating effective radiative forcing from temperature observations

Historical estimates of radiative forcing during the industrial era are usually calculated using climate models from concentrations of forcing agents, such as greenhouse gasses and aerosols. In this study, we diagnose the historical radiative forcing from observed temperature anomalies using a kernel derived from the CMIP5 models (Taylor et al., 2012).This study demonstrates a new application of the temporal kernel method (Larson and Portmann, 2016) to synthesize observations and models.

Good et al. (2011, 2012) showed that the temperature time series ($Y$), and other climate variables, can be approximated by the linear combination of the variable response ($K$) to a step change in forcing ($F_0$) multiplied by the known changes in forcing from all previous years ($\Delta F$). Their step-response model allows simple calculation of temperature time series from radiative forcing time series. Larson and Portmann (2016) demonstrated that the temperature response can be written as a kernel and inverted such that the ERF time series can be calculated from climate model temperature time series using

$$\overrightarrow{\Delta F}/F_0 = K^{-1}\vec{Y} \qquad\qquad (1)$$

The CMIP5 variable responses to the 4xCO$_2$ simulations used to create the kernels are shown in supplementary Figure S1. Thus far, this method has only been applied to model output. Here we apply the temporal kernel method in a novel way to the mean of the GMST anomalies in Figure 1A; NOAA (Vose et al., 2012; Zhang et al., 2017), GISS (GISSTEMP team, 2017; Hansen et al., 2010), Hadcrut (Morice et al., 2012), and BEST (Rohde et al., 2013). The data are normalized to the period between 1880 and 1900. Most of these temperature data sets only go back to 1880, which limits our ability to assess temperature changes before then. This study assumes that the anthropogenic temperature change before 1900 is small compared to the change since and ignoring it does not greatly affect the results. This assumption creates zero historical RF over the reference period (1880-1900), and thus the anthropogenic aerosol which is calculated as a residual offsets the better known GHG and volcanic RFs during that period. For comparison, the mean RF relative to 1765 from Meinshausen et al. (2011) during the reference period is 0.04 W m$_{-2}$. Figure 1B shows the mean and standard deviation of these temperature anomalies.

The historical ERF derived from observed temperature anomalies is shown in Figure 1C and compared with the CMIP5 multi-model mean. The forcing has been smoothed with a 5-year filter to remove noise associated with the kernel method. The two forcing estimates are very similar, suggesting that modeled climate response, as captured by the kernel, works for a range of timescales. There are a couple of decade long temperature anomalies that are not well explained by the CMIP5 models that are reflected in the observationally-derived ERF. These include the anomalously warm period around 1940 and the recent warming slowdown in the 2000s. The difference in the1940s may be due to uncorrected biases in sea surface temperature measurements due to changes in instrumentation or possibly other issues with measurements or emissions inventories during World War II (Thompson et al., 2009). There are several papers discussing the recent warming slowdown and its likely cause, including a changing ocean circulation which redistributed heat into the deep ocean (England et al., 2014; Liu et al. 2016; Meehl et al., 2011) as well as errors in volcanic RF (Santer et al., 2013; Solomon et al., 2011) and other radiative terms (Fyfe et al., 2016; Marotzke and Forster, 2015).

The historical ERF with uncertainty is shown in Figure 2B and compared with the CMIP5 multi-model mean ERF. The uncertainty in the historical ERF comes from two sources, the standard deviation of the temperature datasets and the variation in model kernels. We identify an upper and lower temperature response kernel by using the inter 2/3 of model responses as a range of uncertainty. This range is shown in supplementary Figure S1. To estimate the high (low) end of the uncertainty from these two sources, we calculate the ERF using the GMST plus (minus) one standard deviation and the upper (lower) kernel function.

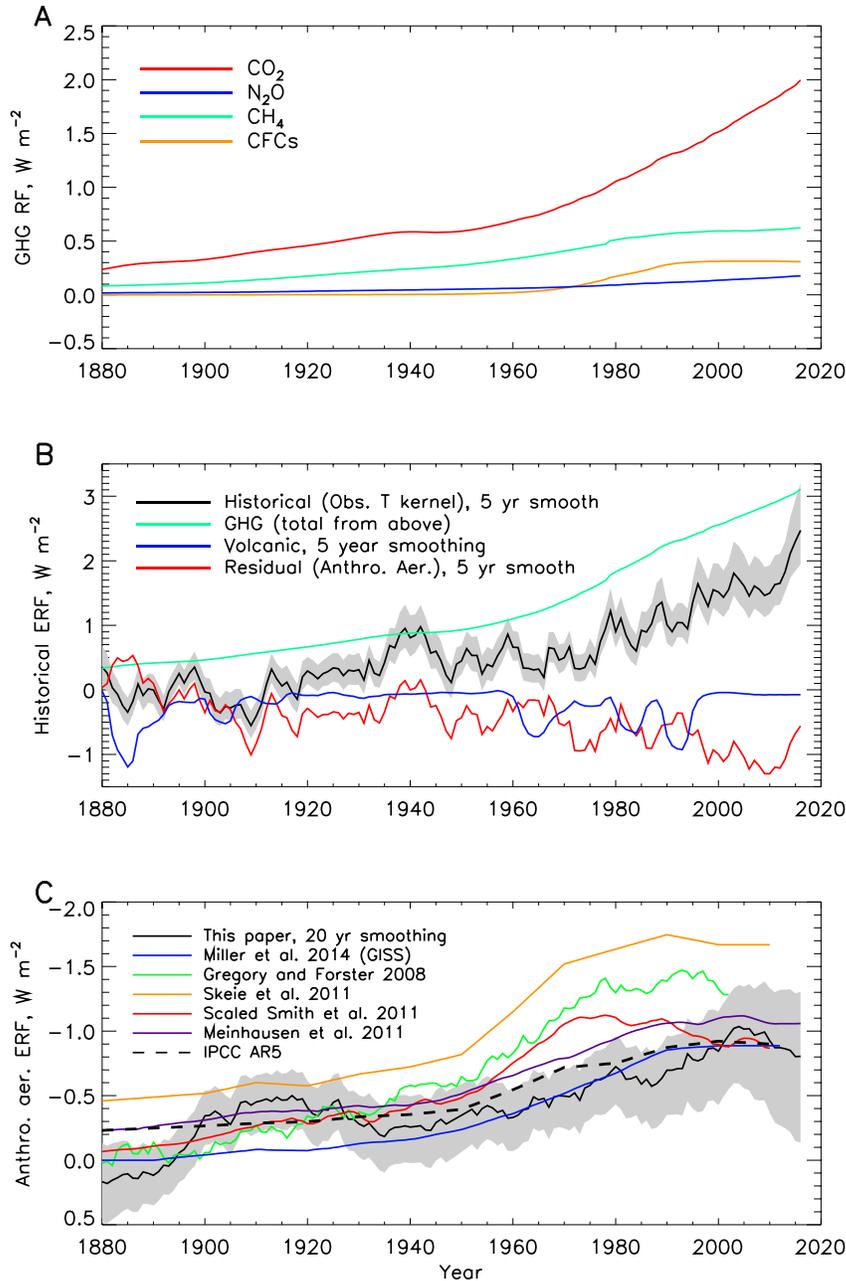

Figure 2. A) The instantaneous radiative forcing of $CO_2$, $CH_4$, $N_2O$ are calculated using the RF formulae from Etminan et al. (2016). RF of the CFCs are calculated using radiative efficiencies (Myhre et al., 2013). B) The anthropogenic aerosol ERF (red line) calculated as a residual from the total instantaneous GHG forcing (green line) and an observationally based estimate of the historical ERF (black line, same as Figure 1c). C) The total anthropogenic aerosol radiative forcing from this study compared with several

other estimates. Grey shading in panels B and C indicate the 1-sigma uncertainty from variation in both the temperature observations and the model responses used in the temporal kernel method calculation.

The anthropogenic aerosol ERF is calculated as a residual from the historical forcing following the method of Murphy et al. (2009). The residual forcing, which is assumed to predominantly represent anthropogenic aerosol, is the difference between the well-mixed GHG forcing and historical forcing shown in Figure 1C. These forcing time series are shown in Figure 2. Figure 2A shows the well-mixed GHG forcing from $CO_2$, $CH_4$, $N_2O$, and the 5 most radiatively active chlorinated species (CFC11, CFC12, CFC113, HCFC22, $CCl_4$), which we label as CFCs. The instantaneous radiative forcing of the $CO_2$, $CH_4$, $N_2O$ are calculated using RF formulae from Etminan et al. (2016). RF of the CFCs are calculated with simple radiative efficiencies (Myhre et al., 2013). We assume that instantaneous and effective GHG radiative forcing are similar enough that differencing the instantaneous GHG forcing and effective historical radiative forcing results in a meaningful approximation of the residual anthropogenic aerosol ERF. This assumption is consistent with Hansen et al. (2005) who report a, likely fortuitous, 1% difference between instantaneous and effective radiative forcing for well mixed greenhouse gases between 1880 and 2000. In general, the difference between instantaneous and effective radiative forcing for several GHGs over a range of perturbations is generally less than 10% (Hansen et al., 2005). Figure 2B shows the cumulative GHG radiative forcing, historical ERF including the uncertainty, the CMIP5 multi-model mean volcanic RF, and the residual aerosol ERF. The anthropogenic aerosol ERF is presently about -1 W/m2, and offset about 1/3 of the GHG radiative forcing over the last decade.

Figure 2C compares the anthropogenic residual aerosol radiative forcing with several estimates from the literature (Gregory and Forster, 2008; Klimont et al., 2011; Miller et al., 2014; Skeie et al., 2011; Smith et al., 2011, Myhre et al. 2013). Our forcing estimate suggests that the early part of the 20th century may have had a substantial amount of aerosol emissions. The CMIP5 kernels require an average of −0.4 W/m2 of ERF from anthropogenic aerosols between 1900 and 1940 to reproduce the temperature observations. However, during the latter part of the 20th century, our estimate is on the lower end and does not require radiative forcing of more than -1W/m2 to match the historical ERF. The IPCC AR5 estimate is within the uncertainty of our residual aerosol ERF estimate throughout the 20th century. Anthropogenic aerosols contribute the largest uncertainty to radiative forcing, with error bars spanning 0 for the 2011 ERF in the IPCC AR5 (Myhre et al., 2013). The uncertainty in our observationally based estimate does not span zero for the entire 20th century except for the period around World War II (WWII), in which global temperature may be biased (Thompson et al., 2009).

There are two major limitations of the current implementation of the temporal kernel method. First, it assumes that each forcing agent has the same efficacy. Ideally, step change experiments for each forcing agent would be carried out, however, this was not done for the full set of CMIP5 models. Therefore, we assume that the climate response to $CO_2$ forcing is the same for all forcing agents This assumption is common when diagnosing radiative forcing in transient simulations (Gregory et al., 2008; Andrews et al., 2012). Multiple kernels should be used in future CMIPS that simulate more step change experiments. This assumption adds a level of uncertainty to the estimated ERF. Hansen et al. (2005) did a very thorough, if dated, quantification of the efficacy of radiative forcing agents. They found that the temperature response was surprising similar across forcing agents, usually within 10% for different forcing agents and definitions of forcing.

Second, this method assumes that the response of the Earth system is constant overtime, and thus all of the change in the temperature record is forced. This includes changes due to natural interannual variability and longer-term variability due to ocean basin oscillations. As mentioned earlier, the warming slowdown during the 2000's was not captured by the CMIP5 models and is likely due to natural variability in the Earth system. Our assumption that anthropogenic aerosols account for the total residual forcing likely overestimates the aerosol forcing between 2000 and 2014. However, the recent GMST increase since 2014 suggests that the long-term behavior of the climate is consistent with model predictions. Two other decades stand out as being influenced by natural variability. The Earth's temperature response to the volcanic aerosol forcing in the late 1800s is weak, either because the efficacy of volcanic forcing is lower than for $CO_2$, or because the volcanic ERF is over estimated. In either case, our residual anthropogenic aerosol forcing is anomalously positive during this time period. The other period with anomalously high residual forcing is during WWII. Despite these limitations the temporal kernel method is a useful tool for diagnosing radiative forcing and adds a novel estimate of the residual anthropogenic aerosol ERF.

## 3 Apportioning current and future climate change by decadal emissions

Not only can we diagnose the historical ERF with temporal kernels, but we can also use them to attribute past emissions to current climate change. Past emissions affect the climate beyond their atmospheric lifetime by affecting the radiative balance of the Earth and the OHC. The contribution to current climate change such as GMST from past emissions is the convolution of the atmospheric lifetime and climate response to the ERF time series from that emission. The kernels also allow us to calculate the relative contribution of each climate forcing agent over time. Attributing current climate change to individual decades of emissions is interesting, as it demonstrates the importance of the lifetime of the forcing agent. This is well established in the literature (Gleckler et al., 2006; Solomon et al., 2010; Zickfeld et al., 2017), however this technique allows us to go beyond idealized simulations and attribute the contribution to current and future temperature change (and OHC) from each past decade in a computationally efficient way.

In this section, we apportion how past decades of emissions of GHGs and aerosol have contributed to current climate change. To do this, we first apportion the contribution from past emissions to current concentrations of GHGs. We track the GHGs that have the largest contribution to radiative forcing and climate change, specifically $CO_2$, $CH_4$, $N_2O$ and the CFCs mentioned above. Starting from observed concentrations (pre-1979 from Meinshausen et al., 2011, 1979-present from NOAA AGGI, Hoffman et al., 2006) we infer emissions timelines of the major GHGs using their decay functions. We use a top down approach so that our emissions estimates are consistent with observed concentration and are self-consistent when using the concentrations to calculate radiative forcing.

We calculate the decay of these gases in the atmosphere using exponentials. $CO_2$ is modeled using a sum of exponential fits based on a multi-model analysis (Joos et al., 2013) with the following coefficients and lifetimes; 21.7% of $CO_2$ having a lifetime of 10,000 years, 22.4% 394.4 years, 28.2% 36.5 years, and 27.6% 4.3 years. Other GHGs ($N_2O$, $CH_4$, CFC11, CFC12, CFC113, HCFC22, and $CCl_4$) are modeled using single exponentials with lifetimes of 121, 12.4, 45, 100, 85, 11.9, and 26 years, respectively (Myhre et al., 2013 [IPCC Table 8.A.1]). Anthropogenic aerosols are included in the forcing calculations and emissions are modeled as emitted in the year the forcing took place with no persistence to later years. Supplementary Figure S2 displays the cumulative emissions and anthropogenic concentration over the historical period.

By tracking the contributed concentration from each year's emissions while considering their atmospheric decay, we can apportion the contribution of each past year's emission to current and future concentrations. Supplementary Figure S3 shows the decadal contributions to present and future concentrations of radiatively important GHGs. The decadal contribution of past emissions to current and future GHG concentrations does not directly correspond to the contributions to current and future temperature (Figure S4) and sea level changes due the memory of the Earth system. The oceans, which warm due to Earth's radiative forcing (Murphy et al., 2009) give the Earth system a memory on timescales relevant to ocean circulation. An increase in OHC due to GHG emissions takes centuries to equilibrate meaning that forcing agents that have been removed from the atmosphere are still contributing to climate change including thermosteric SLR and GMST (Solomon et al., 2010; Zickfeld et al., 2016). Aerosols, for example, have a relatively short atmospheric lifetime, however volcanic eruptions affect OHC for centuries (Gleckler et al., 2006; Gregory et al., 2013).

To calculate how past emissions map to GMST and OHC, we calculate the ERF time series associated with the concentration from emissions in each year. We approximate the ERF ($F_j$ in Eq. 2) of $CO_2$, $CH_4$, and $N_2O$ using the RF formulae from Etminan et al. (2016). This formula produces slightly larger radiative forcing for methane compared with the IPCC AR5 (Myhre et al., 2013; Ch.8 supplement, Table 8.SM.1).The ERF is calculated relative to the preindustrial concentration for the remaining GHG concentration due to emissions in each year. The contributions to the forcing are then scaled to the total forcing for that year based on the fraction of GHG remaining from previous years emissions. This step is necessary because the forcing calculated for a GHG perturbation depends on the background concentration that changes with time. The GHG emissions in each year are given equal weighting (i.e. a 1 ppm perturbation gives the same forcing despite the background concentration in which it was emitted; see below). Radiative forcing for CFCs are calculated using radiative efficiencies of each CFC multiplied by the concentration. The weights are the same as used to calculate the lifetimes. Anthropogenic aerosol ERF is calculated as described above and shown in Figure 2. The equation for calculating the change in forcing for $CO_2$, $CH_4$, and $N_2O$ is

$$\Delta F_{i,j} = \frac{F_j C_{i,j}}{\sum_{i=0}^{j} C_{i,j}} \qquad (2)$$

where i is the year in which a GHG was emitted, j is the year for which the forcing is calculated, $C_{i,j}$ is the concentration of the GHG emitted in year i that is remaining in year j, $F_j$ is the forcing in year j due to the GHG, and $\Delta F_{i,j}$ is the forcing in year j due to the GHG emitted in year i. The sum of $C_{i,j}$ equals the concentration in year j minus the reference period, which is the preindustrial in this case.

As mentioned above, the ERF perturbation in future decades depends on the background concentration for some GHGs ($CO_2$ and $CH_4$). In other words, the RF of an individual $CO_2$ molecule decreases as the $CO_2$ concentration increases. We are left with a choice when attributing current ERF to past decades, whether to calculate an ERF perturbation relative to the $CO_2$ concentration in which the GHG was emitted or relative to the current GHG concentration. Here is a short example illustrating this issue. A kilogram of $CO_2$ is emitted in 1960 when the background concentration is 318 ppm. After 50 years, about half of that kilogram of $CO_2$ remains in the atmosphere. During those 50 years the background concentration went up to 390 ppm. The RF of each molecule of $CO_2$ is about 20% lower in 2010 than in 1960. How should we compare a half kilogram of $CO_2$ emitted in 2010 year with the half kilogram remaining from an emission in 1960? Although the 1960 $CO_2$ was emitted into a lower background, we have decided not to penalize the earlier emissions by calculating their ERF perturbation relative to 318 ppm, but instead, weight all $CO_2$ molecules equally, regardless of when they were emitted. Thus, the forcing perturbation in 2010 from the half kilogram emitted in that year and the half kilogram remaining from 1960 are the same. We consider this the best interpretation and method for attribution of past emissions contributing to current climate change. We note that this choice about ERF attribution may have policy implications if nations decide to impose penalties or regulations for climate change, such as a retroactive carbon tax.

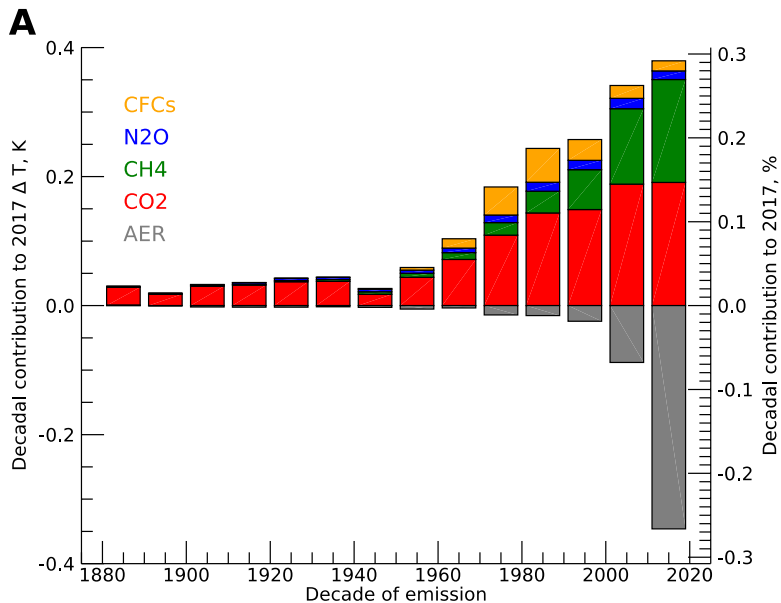

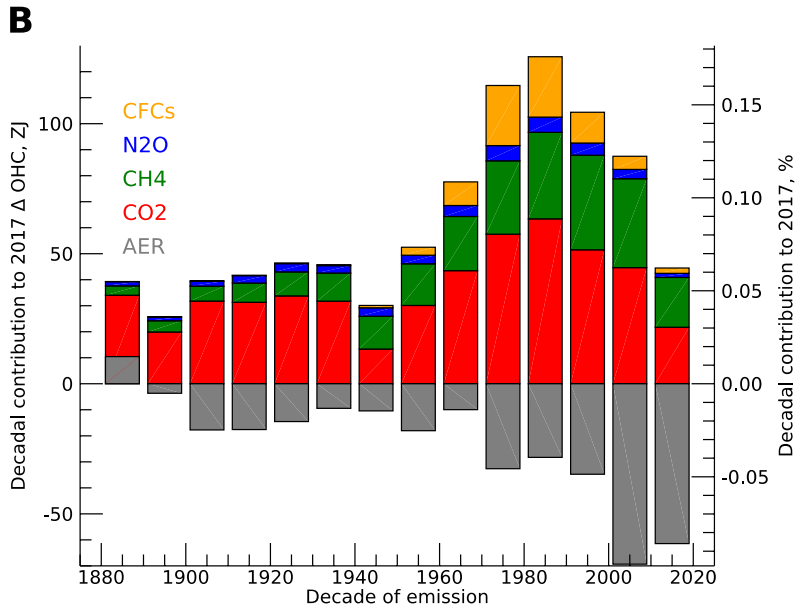

Figure 3. CMIP5 derived decadal contributions to year 2017 (A) GMST and (B) ΔOHC due to anthropogenic GHG and aerosol emissions. The GHG contributions to the 2010's are scaled by 10/8 to make them a full decade of influence. The aerosol contribution was not scaled due to its short lifetime.

The kernel method is applied in the forward direction, i.e. specify forcing to calculate climate variables. In this direction, the kernel method is equivalent to the step response model of Good et al. (2011,2012). Here, we calculate the response of the GMST, OHC, and thermosteric SLR to the ERF calculated above. We bin the yearly variable responses into decadal contributions to current and future climate change. Figure 3 shows the decadal contributions to present day GMST anomaly (A) and ΔOHC (B) since the preindustrial. Decadal contributions to thermosteric SLR look very similar to those of OHC and are not shown in the figure. The current decade's (2010-2017) temperature change and ΔOHC contribution from the GHGs has been scaled by 10/8 to account for the partial years in the decade. The aerosol component was not scaled due to the short lifetimes of aerosols. The linear scaling is a rough estimate of the total decadal contribution from the 2010's, however not scaling the decade gave the misleading impression of a recent reduction in temperature change contribution. The bars in Figure 3 represent the contribution of present day warming (year 2017) due to emissions in that decade. The large anthropogenic aerosol signal in temperature since 2010 is not due to a change in aerosol emissions, but due to the short lifetime of aerosols in the atmosphere. Most of the GMST impact from aerosols occurs in the decade they are emitted. Ocean heat content, however, is impacted more by the integrated radiative forcing and the bar graph reflects that aerosol emissions produce a lasting effect on OHC. This lasting effect is due to the slow response of the oceans to thermally equilibrate. Figure 3 represents the convolution of the lifetime of an emission in the atmosphere with the climate response to that emission's radiative forcing. The impact of aerosols emitted in this decade will quickly diminish due to their short lifetime, whereas $CO_2$ persists in the atmosphere and continue to impact temperature for decades and centuries. Short lived gasses, such as $CH_4$ with a 12-year lifetime, also continue to affect the GMST and OHC after the emission has been removed from the atmosphere. Looking back, past decades of anthropogenic aerosol emissions have a relatively small total effect (~ -0.2 K) on the current temperature anomaly compared with this decade's aerosol emissions (-0.35 K).

The change in OHC is roughly proportional to cumulative ERF, and thus, past emissions have a larger effect on current anomalies as seen in Figure 3B. Emissions of GHGs and aerosols from the 1970s and 1980s have the largest contributions to the current OHC anomaly, despite global $CO_2$-equivalent emissions increasing every decade since, simply because those emissions have remained in the atmosphere longer (Heede, 2014).

Figure 4 shows the relative contribution of past GHG and aerosol emissions to the temperature and OHC anomalies since 1880. The solid line is total temperature anomaly due to anthropogenic forcing agents, which is the observed without the effects of volcanoes. This figure is the cumulative total of the convolution the lifetime of atmospheric emissions with the temperature and OHC responses from the CMIP5 models. $CO_2$ is the largest contributor to temperature change and OHC, followed by anthropogenic aerosols, $CH_4$, CFCs and $N_2O$. Anthropogenic aerosols offset about 1/3 of today's potential GMST and OHC anomalies. Future attempts to reduce aerosol pollution will come at the cost of increasing GMST and OHC.

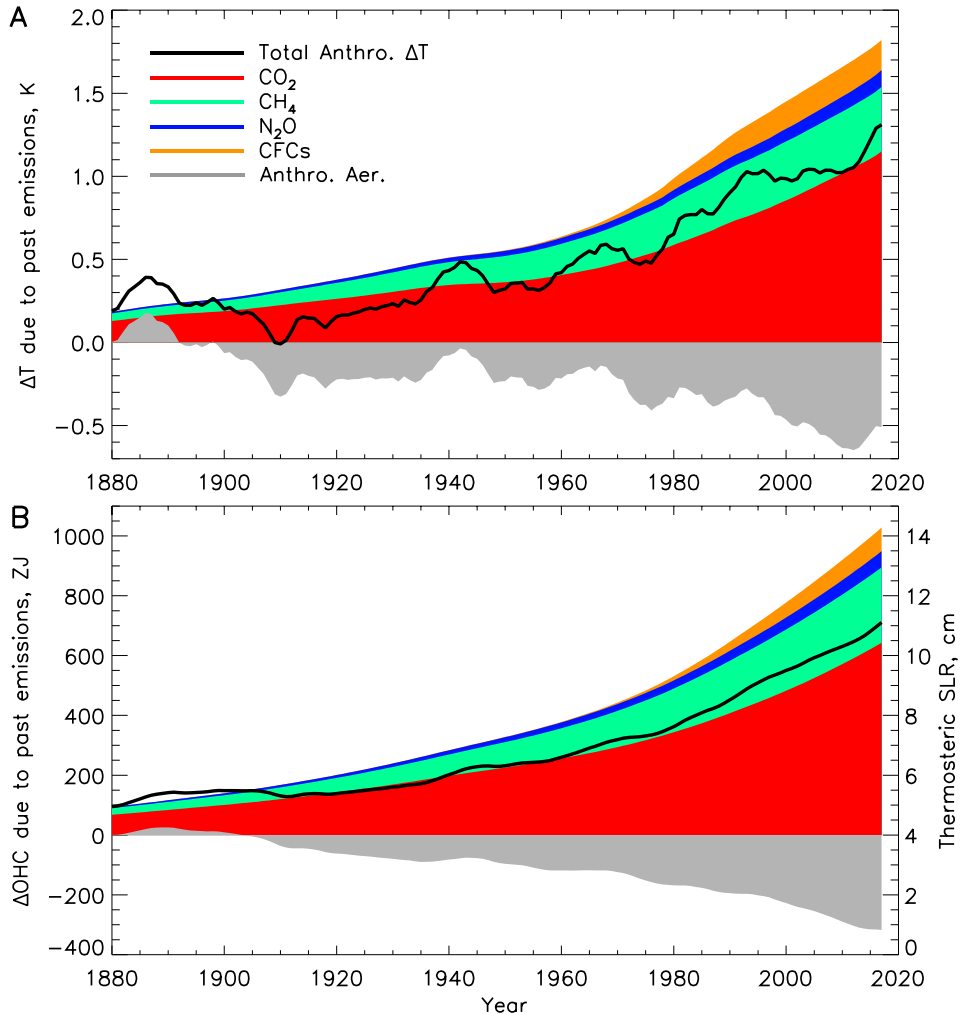

Figure 4. A) GMST anomaly due to past anthropogenic emissions of GHGs and aerosols. B) Global ΔOHC and thermosteric SLR anomaly due to past anthropogenic emissions of GHGs and aerosols. Note that the black lines do not match current ΔT and ΔOHC, as this does not account for the natural component.

## 4 Conclusions

The temporal kernel method (Larson and Portmann, 2016) allows for calculations between ERF and any climate variable time series. This method can be applied in either direction with computational efficiency. We applied the method to the historical GMST record to calculate an observationally based historical ERF. The historical ERF is broadly consistent with purely model-based estimates (Meinshausen et al., 2011; Forster et al., 2013; Larson and Portmann, 2016) and reflects the ERF that the average climate model would need to recreate the historical temperature record. Although this method relies heavily on model-derived kernels, it leverages the best historical climate record available to estimate the forcing.

Anthropogenic aerosol ERF is calculated as a residual from the historical ERF and the relatively well known GHG radiative forcing. This ERF suggests that anthropogenic aerosol radiative forcing was several tenths of a W $m_2$ in the early part of the $20_{th}$ century, consistent with a large aerosol indirect effect. The aerosol ERF is consistent with the estimate from the IPCC AR5 and is approximately -1 W/$m_2$ in the 2010s. Furthermore, the uncertainty in this estimate is less than in the AR5. One limitation of this method is that natural variability in the temperature signal gets mapped onto the residual aerosol forcing. For instance, the temperature slowdown in the 2000's manifests as large aerosol forcing, when it is likely due to a change in ocean mixing (England et al., 2014; Liu et al., 2016; Meehl et al., 2011).

Understanding how GHG and aerosol emissions contributed to climate change requires accounting for the lifetime of the emissions as well as how the change in forcing from the emitted species is manifest as temperature change or OHC. The current GMST anomaly has large contributions from $CO_2$ emissions over the past century and more recent other GHGs. However, much of the current potential warming is offset by more recent aerosol emissions over the past two decades. By contrast, OHC and thermosteric SLR are affected by cumulative radiative forcing and the current anomalies have large contributions from both GHGs and aerosols in past decades.

This difference of timescale has important policy implications. For example, if GHG emissions were reduced to zero, OHC and SLR will still be higher in 2100 than at present, but the GMST would be slightly lower. Thus, SLR is irreversible over the next century without active carbon sequestration or climate modification.

The utility of kernels to calculate between effective forcing and climate variables in either direction is demonstrated by our ability to apportion current and future climate change by decadal emissions. The temporal kernels are also computationally efficient, especially compared to climate models. This paper also demonstrates a method in which one could specify future emissions and calculate the associated ERF, GMST, and SLR. These estimates would reflect the average CMIP5 climate model responses to the specified emissions. These kernels have wide application in both diagnosing model simulations and calculating radiative forcing from observations, both of which can be used to inform policy decisions.

**Acknowledgements**
The CMIP5 temperature response kernels to the 4xCO2 forcing experiments are publicly available online (https://github.com/larsonej/CMIP5_kernels). The authors would like to thank the respective agencies that collected and compiled the surface air temperature data. SS gratefully acknowledges partial support under NSF grant 1848863. The temperature time series data can be obtained from the respective agencies: GISS (https://data.giss.nasa.gov/gistemp/), NOAA (https://www.esrl.noaa.gov/gmd/aggi/aggi.html), Hadley

(https://crudata.uea.ac.uk/cru/data/temperature/), Berkeley
(http://berkeleyearth.org/data/).